\documentclass{article}

\usepackage[nonatbib,preprint]{neurips_2021}
\usepackage{amsmath}
\usepackage{graphicx}
\usepackage{hyperref}
\usepackage[square,sort&compress,comma,numbers]{natbib}
\usepackage{xfrac}

\usepackage[utf8]{inputenc} 
\usepackage[T1]{fontenc}    
\usepackage{hyperref}       
\usepackage{url}            
\usepackage{booktabs}       
\usepackage{amsfonts}       
\usepackage{nicefrac}       
\usepackage{microtype}      
\usepackage{xcolor}         

\usepackage{float}
\definecolor{antiquefuchsia}{rgb}{0.57, 0.36, 0.51}
\definecolor{darkred}{RGB}{135, 33, 9}
\definecolor{cadetblue}{rgb}{0., 0.55, 0.55}
\definecolor{darkgrey}{rgb}{0.35, 0.35, 0.35}
\definecolor{maroon}{rgb}{0.76, 0.0, 0.08}
\hypersetup{
    colorlinks = true,
    linkbordercolor = {white},
    citecolor = cadetblue,
    linkcolor = cadetblue,
    menucolor = cadetblue,
    urlcolor = darkred
}
\usepackage{color}
\usepackage{wrapfig}

\usepackage[linesnumbered,ruled,vlined]{algorithm2e}
\usepackage{xparse}

\newif\iffirstitem
\firstitemtrue

\SetCommentSty{mycommfont}
\newcommand{\nosemic}{\renewcommand{\@endalgocfline}{\relax}}
\newcommand{\dosemic}{\renewcommand{\@endalgocfline}{\algocf@endline}}
\let\oldnl\nl
\newcommand{\nonl}{\renewcommand{\nl}{\let\nl\oldnl}}

\renewcommand\theequation{{\color{magenta}\arabic{equation}}}

\newcommand\inlineeqno{\stepcounter{equation}\ (\theequation)}

\renewcommand{\eqref}[1]{({\color{gray}\rref{#1}})}

\AtBeginDocument{\newcommand{\rref}[1]{{\color{gray}\ref{#1}}}}

\title{Deterministic particle flows for constraining SDEs}

\author{%
  Dimitra Maoutsa \\
  Technical University of Berlin\\
  \&  University of Potsdam\\
  \texttt{\color{darkgrey}{dimitra.maoutsa@tu-berlin.de}} \\
   \And
  Manfred Opper \\
    Technical University of Berlin \\
    \&  University of Birmingham \\
    \texttt{\color{darkgrey}{manfred.opper@tu-berlin.de}} \\
}

\begin{document}

\maketitle
\begin{abstract}
  Devising optimal interventions for diffusive systems often requires the solution of the Hamilton-Jacobi-Bellman ({HJB}) equation, a nonlinear backward partial differential equation ({PDE}), that is, in general, nontrivial to solve. Existing control methods either tackle the HJB directly with grid-based PDE solvers, or resort to iterative stochastic path sampling to obtain the necessary controls.
  Here, we present a framework that interpolates between these two approaches.
  By reformulating the optimal interventions in terms of logarithmic gradients (\emph{scores}) of two forward probability flows, and by employing deterministic particle methods for solving Fokker-Planck equations, we introduce a novel fully \emph{deterministic} framework that computes the required optimal interventions in \emph{one shot}. 
\end{abstract}

\section{Introduction}

\paragraph{Constrained diffusions and optimal control.} Consider the problem of imposing constraints $\mathcal{C}$ to the state of a stochastic system, whose \emph{unconstrained} dynamics are described by a stochastic differential equation (SDE) \begin{equation}\label{eq:free_SDE} 
  dX_t = f(X_t,t) dt  + \sigma dW_t,
\end{equation}
with drift $f(x,t)\in \mathcal{R}^d\times \mathcal{R} \rightarrow \mathcal{R}^d$ and diffusion coefficient $\sigma \in \mathcal{R}$.\footnote{For brevity, we restrict ourselves to state- and time- independent scalar diffusions, but the framework easily generalises to time-dependent multiplicative noise settings with non-isotropic diffusion functions.} For a time interval $[0,\,T]$, the constraints $\mathcal{C}$ may pertain either the transient state of the system through a path-penalising function $U(x,t)\in \mathcal{R}^d\times \mathcal{R} \rightarrow \mathcal{R}$ for $t \leq T$, or its terminal state $X_T$ through the function $\chi(x)\in \mathcal{R}^d \rightarrow \mathcal{R}$. 

One way to obtain the path probability measure $\mathbb{Q}$ of the \emph{constrained} process is by {\em reweighting} paths
$X_{0:T}$ generated from Eq.(\ref{eq:free_SDE}) over the interval $[0,\, T]$.  
Path weights result from the {\em Radon--Nikodym derivative} with respect to the path measure of the unconstrained process $\mathbb{P}_f$
\begin{equation} \small
\frac{d\mathbb{Q}}{d\mathbb{P}_f} (X_{0:T}) = \frac{\chi(X_T)}{Z} \exp\left[- \int_0^T U(X_t,t) dt \right],
\label{Reweight}
\end{equation}
where $Z$ is a normalising constant. 

How shall we modify the system of Eq.\eqref{eq:free_SDE} to incorporate the desired constraints $\mathcal{C}$ into its dynamics, while also 
minimising the relative entropy between the path distributions of the constrained and unconstrained processes? 

Problems of this form appear often in physics, biology, and engineering, and are relevant for calculation of rare event probabilities~\cite{hartmann2012efficient,chetrite2015variational}, latent state estimation of partially observed systems~\cite{kim2020optimal,todorov2009efficient}, or for precise manipulation of stochastic systems to target states~\cite{kappen2005linear,wells2015control} with applications on artificial selection~\cite{nourmohammad2021optimal,evolver}, motor control~\cite{kao2021optimal}, epidemiology, and more~\cite{iolov2014stochastic,scott2004optimal,bernton2019schr,vargas2021solving,exarchos2018stochastic,song2020score}.
Yet, although stochastic optimal control problems are prevalent in most scientific fields, their numerical solution remains computationally demanding for most practical problems.

The constrained process, defined by the weight of Eq.(\ref{Reweight}), can be also expressed as a diffusion process with the same diffusion $\sigma$, but with a modified time-dependent drift $g(x,t)$~\cite{girsanov1960transforming,oksendal2003stochastic}. The computation of this 
drift adjustment or {\em control} $u(x,t) \doteq g(x,t) - f(x,t)$ 
amounts to solving a
{\em stochastic control} problem to obtain the {\em optimal control}
$u^*(x,t)$ that minimises the expected cost 
\begin{align} \small
\mathcal{J}(x,t) \doteq \min_u \mathbb{E}_{\mathbb{P}_g}\left[ \int_0^T \left( \frac{1}{2 \sigma^2} \| u(X_t,t)\|^2  + 
 U(X_t ,t) \right) dt - \ln \chi(X_T) \right].
 \label{opt_control}
\end{align}
The expectation $\mathbb{E}_{\mathbb{P}_g}$ is over paths induced by the {\em constrained} SDE  $ \; dX_t = g(X_t,t) dt  + \sigma dW_t.\;\;\; \label{eq:controlled_SDE}  \inlineeqno$\\
 This control setting is known as {\em Path Integral-} or {\em Kullback-Leibler-control} (\textbf{PI/KL-control})\cite{todorov2009efficient,kappen2005linear,theodorou2012relative, kappen2012optimal}. Finding the exact optimal 
interventions for general stochastic control problems amounts to solving 
the HJB equation~\cite{bellman1956dynamic}, a computationally demanding {\em nonlinear} PDE. 
Yet, for the PI-control problem, the logarithmic Hopf-Cole transformation~\cite{fleming1977exit}, ie. ${\mathcal{J}(x,t) = -\log( \phi_t(x))}$, linearises the HJB equation~\cite{kappen2005linear}, and the optimally adjusted drift becomes 
\noindent\begin{minipage}{.5\linewidth}
\begin{equation}
g(x,t)  = f(x,t) +  \sigma^2 \nabla \ln \phi_t(x), 
\label{newdrift}
\end{equation}
\end{minipage}%
\begin{minipage}{.5\linewidth}
\begin{equation}
 \frac{\partial \phi_t(x)}{\partial t} + {\cal{L}}^{\dagger}_f \phi_t(x) - U(x,t) \phi_t(x) = 0  ,
\label{backward_equation}
\end{equation}
\end{minipage}
where $\phi_t(x)$ is a solution to the backward PDE of Eq.\eqref{backward_equation} with terminal condition $\phi_T(x) = \chi(x) $, and $\mathcal{L}^{\dagger}_f \phi(x)\doteq f(x,t)\nabla\cdot  \phi (x) + \frac{\sigma^2}{2} \nabla^2 \phi(x)$ denotes the 
\emph{adjoint} Fokker--Planck operator acting on $\phi(x)$.

The PDE of Eq.\eqref{backward_equation} is often treated either with grid based solvers~\cite{garcke2017suboptimal,annunziato2013fokker}, or with iterative stochastic path sampling schemes ~\cite{kappen2012optimal,kappen2005linear,kappen2016adaptive,zhang2014applications}.  
Both approaches suffer, in general, from high computational complexity with
increasing system dimension. (However note recent neural network advances towards this direction~\cite{macris2020solving, li2020fourier}.)


\section{Method}
\paragraph{Constrained diffusion densities from backward smoothing.}
 To avoid directly solving the backward PDE \textcolor{gray}{(Eq.~\eqref{backward_equation})}, we view the {\em marginal density} $q_t(x)$ of the constrained process as the {\em smoothing} density in an inference setting. By considering Eq.\eqref{Reweight} as a likelihood function, 
and treating the constraints $U(x,t)$ and $\chi(x)$ as continuous time observations of the process $X_t$, the path measure $\mathbb{Q}$, i.e. the product of the {\em a priori} distribution 
$\mathbb{P}_f$ with the likelihood, can be interpreted as the 
posterior distribution over paths that account for the observations as soft constraints. Thus,
in a similar vein to the {\em forward--backward} smoothing algorithms
for hidden Markov models~\cite{ephraim2002hidden,baum1970maximization}, we factorise the marginal density into two terms that account for past and future constraints separately
 \begin{equation} \label{eq:factorised}
q_t(x) \propto \rho_t(x)  \phi_t(x).  
\end{equation}
The density $\rho_t(x)$ satisfies the forward \emph{filtering} equation with initial condition $\rho_0(x)$ \textcolor{gray}{(Eq.\eqref{FPE2})},
while the marginal constrained density $q_t(x)$ fulfils a Fokker--Planck equation \textcolor{gray}{(Eq.\eqref{FPE_control})} 
\noindent\begin{minipage}{.5\linewidth}
\begin{equation}
\frac{\partial \rho_t(x)}{\partial t} = {\cal{L}}_f \rho_t(x) - U(x,t) \rho_t(x), 
\label{FPE2} 
\end{equation}
\end{minipage}%
\begin{minipage}{.5\linewidth}
\begin{equation}
  \frac{\partial q_t(x)}{\partial t} = {\cal{L}}_g q_t(x) .
\label{FPE_control} 
\end{equation}
\end{minipage}
Here the Fokker--Planck operator ${\cal{L}}_g$ is defined for the process with optimal drift $g(x,t)$ \textcolor{gray}{(Eq.\eqref{newdrift})}.
By replacing $\phi_t(x)$ with $q_t(x)/\rho_t(x)$ in 
Eq.\eqref{newdrift}, we obtain a new representation of 
the optimal drift in terms of the logarithmic gradients (\emph{score functions}) of two forward probability flows, $q_t(x)$ and $\rho_t(x)$
\begin{equation} \small
g(x,t)  = f(x,t) +  \sigma^2 \Big(\nabla \ln q_t(x) - \nabla \ln \rho_t(x)\Big) .
\label{newdrift2}
\end{equation}
This formulation still requires knowledge of the unknown $\nabla \ln q_t(x)$. Yet, by
inserting Eq.\eqref{newdrift2} into Eq.\eqref{FPE_control}, and introducing a time-reversion through the variable $\tau = T - t$, we obtain a Fokker--Planck equation for the flow $\tilde{q}_\tau(x) = q_{T-\tau} (x)$
\begin{equation}\label{Fokker_bridge3} \small
\frac{\partial \tilde{q}_{\tau}(x)}{\partial \tau} = 
-\nabla\cdot \Big(\big(\sigma^2\nabla \ln  \rho_{T-\tau} (x)  - f(x, T-\tau)\big) \tilde{q}_{\tau} (x)\Big)
+ \frac{\sigma^2}{2} \nabla^2 \tilde{q}_{\tau} (x),
\end{equation}
that depends only on the time-reversed forward flow $\rho_t(x)$, with $\tilde{q}_{0}  \propto \rho_T(x) \chi(x)$.
Thus, for the exact computation of the optimal controls ${ u^*(x,t)=\sigma^2 \big(\nabla \ln \tilde{q}_{T-t}(x) - \nabla \ln \rho_t(x)\big)}$, we require the logarithmic gradients of two forward probability flows $\tilde{q}_{T-t}(x)$ and $\rho_t(x)$.
\paragraph{Deterministic particle dynamics.}
To sample the two forward flows $\rho_t(x)$ and $\tilde{q}_{T-t}(x)$ \textcolor{gray}{(Eq.\eqref{FPE2} and Eq.\eqref{FPE_control})} we employ a recent deterministic particle framework for solving Fokker--Planck equations~\cite{Maoutsa_2020}, modified to fit our purposes. We approximate $\rho_t(x)$ with the empirical distribution
$\hat{\rho}_t(x) \approx \frac{1}{N}\sum_{i=1}^N \delta \left(x - X^{(i)}_t\right)$
constructed from an ensemble of $N$ "particles" $X^{(i)}_t$. 
\begin{wrapfigure}{l}{0.45\textwidth}
\hspace{-20pt}
\includegraphics[scale=0.225]{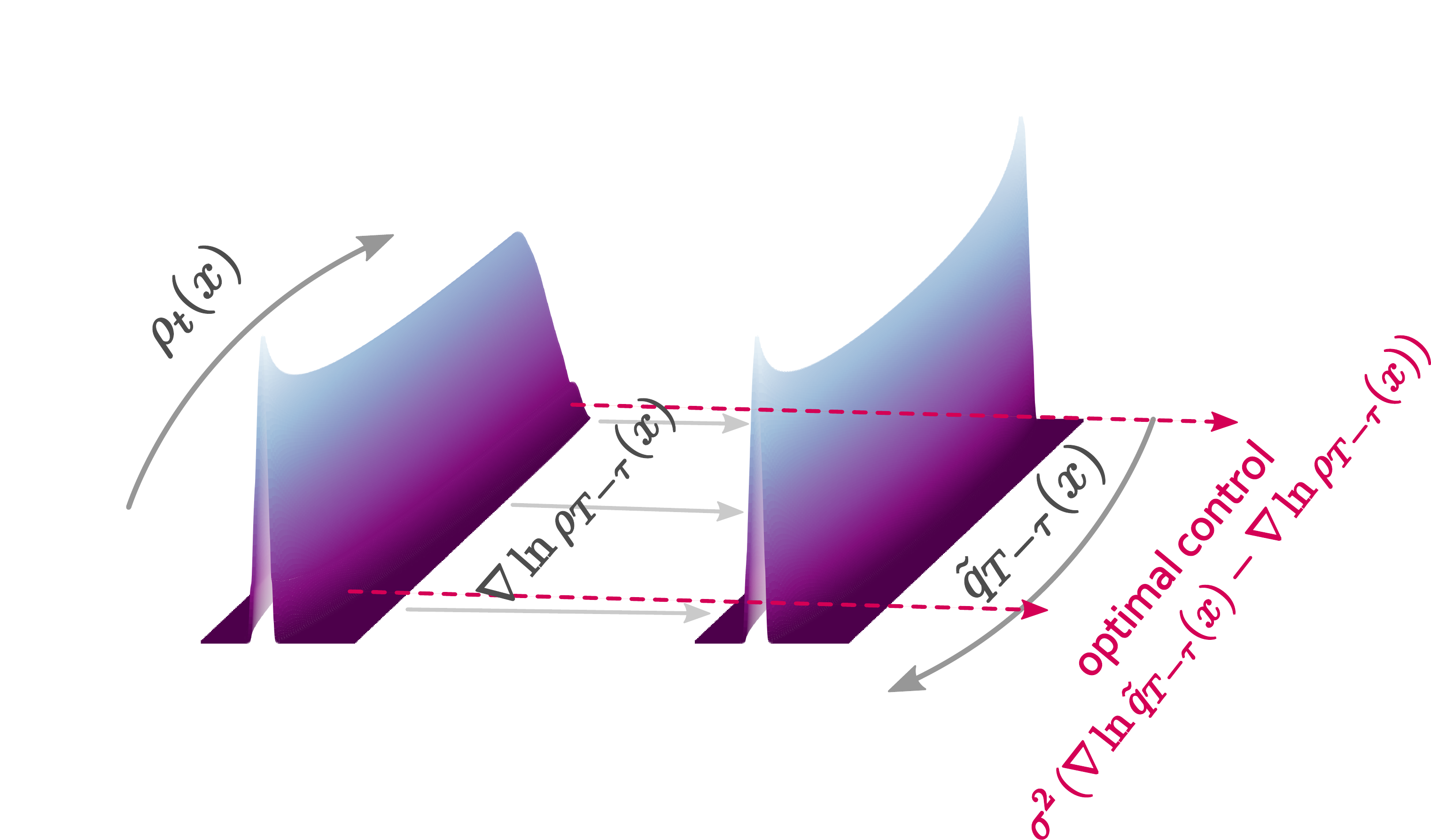}
\caption{\textbf{Schematic of the proposed control framework.}}

\end{wrapfigure}
For flows \emph{without} path costs $\left(U(x,t)\equiv 0\right)$, we express the particle dynamics as a system of ordinary differential equations (ODEs)~\cite{Maoutsa_2020}
\begin{equation} \label{eq:deter_dynamics} 
   d X^{(i)}_t = f(X^{(i)}_t,t)\,dt - \sfrac{\sigma^2}{2}  \hat{\nabla} \ln \hat{\rho}_t(X^{(i)}_t)\,dt,
\end{equation}
where $ \hat{\nabla} \ln \hat{\rho}_t(X^{(i)}_t)$ denotes the estimated score function of the empirical distribution $\hat{\rho}_t(x)$.

For flows \emph{with} path costs $\left(U(x,t)\neq 0\right)$, the flow dynamics in terms of operator exponentials\footnote{In the second equality,  we considered that for small $\delta t$ the commutator of the two operators 
${\cal{L}}_f$ and $U(x,t)$ is negligible.} reads
\begin{align} 
\rho_{ t + \delta t}(x)& = e^{\delta t ({\cal{L}}_f  - U(x,t))} \rho_t(x) \label{two_stage} \\
&=  e^{-  \delta t U(x,t) } e^{\delta t {\cal{L}}_f} \rho_t(x) + \mathcal{O}((\delta t)^2).\nonumber
\end{align}
We interpret Eq.\eqref{two_stage} as the concatenation of two processes:
a density propagation by the uncontrolled Fokker--Planck equation, 
followed by a multiplication by a factor $e^{- \delta t U(x,t)}$. 
To simulate this two-stage process for a time interval $\delta t$,  we first 
evolve the particles following Eq.\eqref{eq:deter_dynamics} to auxiliary positions $Y^{(i)}_t$ and 
assign to each particle $i$ a {\em weight}
$\Omega_i(t) \propto e^{-\delta t U(Y^{(i)}_t,t)}.$
To transform the weighted particles to unweighted ones, we employ the {\em ensemble transform particle filter}~\cite{reich2013nonparametric}. This method provides an optimal transport map that \emph{deterministically} transforms an ensemble of weighted particles into  an ensemble of uniformly weighted ones by maximising the covariance between the two ensembles.
\paragraph{Sparse kernel score function estimator.}
To estimate the scores of the flows $\rho_t(x)$ and $\tilde{q}_t(x)$ for the particle evolution \textcolor{gray}{(Eq.\eqref{eq:deter_dynamics})} and the estimation of optimal controls $u^*(x,t)$, we employ a \emph{sparse kernel score function estimator}~\cite{Maoutsa_2020}. 
More precisely, we obtain each dimensional component $a \in [1,\dots, d]$ of $\nabla \ln\rho(x)$ from the solution of the \emph{variational} problem of minimising the functional $\mathcal{I}_\alpha [h, \rho] $
\begin{equation}  \vspace{-0pt}
{\partial_\alpha} \ln \rho(x) = 
\arg\min_h\; {\cal{I}}_\alpha [h, \rho]{(x)} = 
\arg\min_h\; \int \rho(x) \left( 2 \partial_\alpha h(x) +h^2(x) \right) dx.
\label{esti1}
\end{equation}
To regularise this optimisation we assume that $h$ belongs to a Reproducing Kernel Hilbert Space associated with a radial basis function kernel, and employ a \emph{sparse kernel approximation} by expressing $h$ as a linear combination of the kernel evaluated at  $M \ll N$ {\em inducing points} $Z_i$,
$h(x) = \sum_{i=1}^M b_i K(Z_i, x)$. (See Appendix\ref{app:esti} for the exact formulation of the estimator.)
\begin{figure}[ht!]\label{result}
\vspace{-20pt}
\hspace{-20pt}
\includegraphics[scale=0.375]{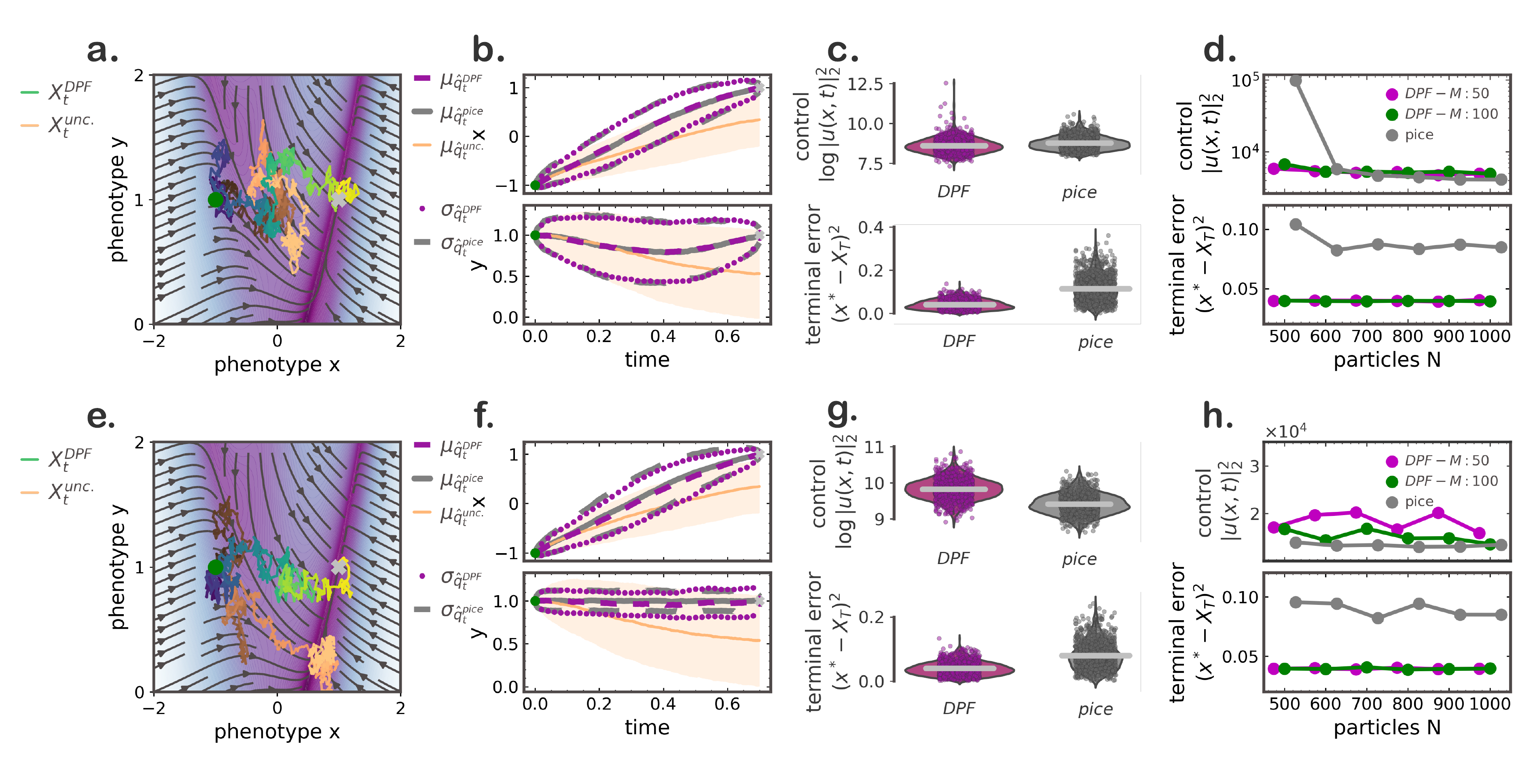} \vspace{-10pt}
\caption{\textbf{Deterministic particle flow (DPF) control provides optimal interventions to drive the system to target state ({grey cross}).} \textbf{(a.)} Example controlled trajectory ({blue-yellow}) successfully reaches target, while an uncontrolled one ({orange}) remains in the vicinity of initial state for the same time interval. \textbf{(b.)} Agreement between summary statistics of marginal densities estimated from $1000$ trajectories controlled by DPF (purple) and PICE (grey) (transient mean $\mu_{\hat{q}_t}$ and standard deviation $\sigma_{\hat{q}_t}$). Orange indicates mean and standard deviation of $1000$ uncontrolled trajectories. (Used $N=400$ particles for DPF, and $N=500$ for PICE to obtain the optimal controls.) \textbf{(c.)} Comparison of \textbf{(upper)} (logarithmic) control energy, and \textbf{(lower)} deviation of terminal state from target for each controlled trajectory (dots) with interventions computed according to DPF (magenta) and PICE (grey). Light grey lines identify the mean of each quantity over the $1000$ trajectories. \textbf{(d.)}\textbf{(upper)} Control energy, and \textbf{(lower)} terminal error for increasing particle number $N$. (inducing point number for DPF magenta: $M=50$, green: $M=100$ ). Grey line indicates the performance of PICE in the same setting. \textbf{(e.-h.)} Same as (a.-d.) with additional path constraint $U((x,y),t)= 10^3  (y-1)^2$.
  }   \vspace{-10pt}
\end{figure}

\section{Numerical Experiments}
We employed our method (deterministic particle flow control-\textbf{DPF}) on a model that can be thought of as describing the \emph{mean} phenotype $(x,y)$ of a population
evolving on a phenotypic landscape $F$ under adaptive pressures $f([x,y],t)=\nabla F(x,y)=\nabla ((1-x)^2 + (y - x^2)^2)$ and genetic drift represented by white noise~\cite{nourmohammad2021optimal}.(See Appendix for more biological relevance.)

Starting from initial state $\textbf{x}_0=(-1,1)$, we evaluate our framework on two scenarios: one with only terminal constraints $\chi(\mathbf{x})= \delta(\mathbf{x}-\mathbf{x}^*)$, with $\mathbf{x}^*=(1,1)$ (Figure~\ref{result}(a.-d.)), and one with the same terminal constraints coupled with a path cost that limits fluctuations along the $y$ axis (Figure~\ref{result}(e.-h.)). In both settings, we benchmark our method against the path integral cross entropy method (\textbf{PICE})~\cite{kappen2016adaptive}, by comparing summary statistics, control costs $\|u(\mathbf{x},t)\|^2_2$, and deviations from target $\|X_T - \mathbf{x}^*\|^2$ of $1000$ independent trajectories controlled by each framework (purple:DPF, grey:PICE).  

Our method successfully controlled the system  
towards the predefined target ($\mathbf{x}^*$-grey cross)  (Figure~\ref{result} a.,e.), and showed complete agreement with PICE in terms of the transient mean and standard deviation of the marginal densities $q_t(x)$ captured by the $1000$ trajectories controlled with each method.
Comparing the control effort characterising the optimality of the interventions (Figure~\ref{result} c.,g.), both methods dissipated comparable energy with DPF showing slightly larger variance among individual trajectories. Nevertheless, by examining the terminal errors, DPF was consistently more accurate and precise in reaching the target. 
Comparing the performance of both methods for increasing particle number employed for obtaining the controls, our approach delivered more efficient controlled from PICE for small number of particles ($N=500$), while both methods were comparable for increasing particle number (Figure~\ref{result} d.).
These results suggest that our method delivers equally optimal controls with PICE in one-shot, while it is also relatively more accurate in reaching the targets. 

\section{Conclusions}
Forward-backward algorithms for smoothing densities have been extensively used in hidden Markov models. Here by relying on the duality between inference and control~\cite{todorov2008general,kappen2012optimal,levine2018reinforcement,attias2003planning}, we borrowed ideas from the inference literature to derive non-iterative sampling schemes for constraining diffusive systems. By employing score function estimation, and recent deterministic approaches for solving Fokker--Planck equations, we proposed a non-iterative stochastic control framework that relies solely on deterministic particle dynamics. Our method interpolates between classical space-discretising PDE solutions that are inherently non-iterative, and stochastic Monte Carlo 'particle' methods that rely on the  Feynaman-Kac formula to obtain PDE solutions through sample paths.

The major limitation 
in applying our approach more broadly in higher dimensional systems is the curse of
dimensionality. The number of particles required to provide enough evidence for accurate score estimation increases with system dimension, and more advanced methods of model reduction shall be combined with the present work. 
A further computational bottleneck when path constraints are pertinent is the computational complexity of ensemble transform particle filter algorithm ($\mathcal{O}(N^3\log N)$~\cite{bertsimas1997introduction}).Yet, there is room for improvement here by applying entropy regularised approaches for particle reweighting~\cite{corenflos2021differentiable}.

\newpage

\if false

\section{Broader Impact}

We provided a deterministic particle approach for controlling stochastic systems that is valuable and useful for computational and applied physicists, applied mathematician, and biophysicists. 
We do not foresee direct social impact. However we acknowledge that stochastic control systems are used for military purposes and financial engineering, that under the wrong circumstances may lead to unfavourable outcomes.

On a more positive note, diffusive systems are prevalent in several scientific fields, such as parts of physics, biology, finance, and ecology.
We foresee that this work may benefit these disciplines either by providing a easy to implement stochastic control tool, or by enabling faster simulation based inference~\cite{brehmer2021simulation,cranmer2020frontier} of sparsely observed systems as a data augmentation tool. 

 In particular, for the field of systems biology, we expect our method to be useful for continuous culturing devices like the  
eVOLVER~\cite{wong2018precise} system, an open-source automated culturing platform, that allows for high throughput experiments for studying evolutionary processes.


\fi

\medskip

{
\small


}

\if false
\section*{Checklist}

\begin{enumerate}

\item For all authors...
\begin{enumerate}
  \item Do the main claims made in the abstract and introduction accurately reflect the paper's contributions and scope?
    \answerYes{}
  \item Did you describe the limitations of your work?
    \answerYes{}
  \item Did you discuss any potential negative societal impacts of your work?
    \answerYes{\textcolor{darkgrey}{Although we do not foresee direct societal impact, since stochastic control algorithms may be used for malicious military and financial engineering purposes, we acknowledge this possibility}}
  \item Have you read the ethics review guidelines and ensured that your paper conforms to them?
    \answerYes{}
\end{enumerate}

\item If you are including theoretical results...
\begin{enumerate}
  \item Did you state the full set of assumptions of all theoretical results?
    \answerYes{\textcolor{darkgrey}{The main assumptions of our approach are that the dynamics of the system under consideration may be captured by an SDE which we consider \emph{known}, and that the control costs arise in a quadratic form in the cost functional, i.e. the cost functional relevant for the system under consideration has the form of Eq.~\eqref{opt_control}.} }
	\item Did you include complete proofs of all theoretical results?
    \answerYes{\textcolor{darkgrey}{The main results are Eq.\eqref{Fokker_bridge3} and evolution of the probability flows with deterministic particle dynamics. We indicate in the main text how to arrive to Eq.\eqref{Fokker_bridge3}, and in the paragraph \textbf{Deterministic particle dynamics} we describe our approach for representing the probability flows $\rho_t(x)$ and $\tilde{q}_t(x)$ with deterministic particle dynamics.} }
\end{enumerate}

\item If you ran experiments...
\begin{enumerate}
  \item Did you include the code, data, and instructions needed to reproduce the main experimental results (either in the supplemental material or as a URL)?
    \answerYes{\textcolor{darkgrey}{We will provide the code upon acceptance. Nevertheless in the main text we give all the parameters used for the numerical experiments (mainly in the Figure 2 caption), and additionally provide in the appendix the exact formulation of the score function estimator (see Appendix~\ref{app:esti}) and a pseudo-code outlining the main framework (see Appendix~\ref{implementation}).}}
  \item Did you specify all the training details (e.g., data splits, hyperparameters, how they were chosen)?
    \answerYes{\textcolor{darkgrey}{The "training" in our setting is the evaluation of the kernels. We provide in the appendix details on the kernels, hyperparameters, and precise form of the score estimator.}}
	\item Did you report error bars (e.g., with respect to the random seed after running experiments multiple times)?
    \answerYes{In Figure~\ref{result}.}
	\item Did you include the total amount of compute and the type of resources used (e.g., type of GPUs, internal cluster, or cloud provider)?
    \answerNo{We didn't use any GPU, cluster etc. Only a plain personal laptop.}
\end{enumerate}

\item If you are using existing assets (e.g., code, data, models) or curating/releasing new assets...
\begin{enumerate}
  \item If your work uses existing assets, did you cite the creators?
    \answerYes{\textcolor{darkgrey}{In the Appendix, we do cite FastEMD, employed for solving the optimal transport problem in the ensemble transform particle filter, and Seaborn used for data visualisation. We benchmarked our method against the Path Integral Cross Entropy framework (PICE) that we implemented alone and cite in the main text.}}
  \item Did you mention the license of the assets?
    \answerNA{}
  \item Did you include any new assets either in the supplemental material or as a URL?
    \answerNA{}
  \item Did you discuss whether and how consent was obtained from people whose data you're using/curating?
    \answerNA{}
  \item Did you discuss whether the data you are using/curating contains personally identifiable information or offensive content?
    \answerNA{}
\end{enumerate}

\item If you used crowdsourcing or conducted research with human subjects...
\begin{enumerate}
  \item Did you include the full text of instructions given to participants and screenshots, if applicable?
    \answerNA{}
  \item Did you describe any potential participant risks, with links to Institutional Review Board (IRB) approvals, if applicable?
    \answerNA{}
  \item Did you include the estimated hourly wage paid to participants and the total amount spent on participant compensation?
    \answerNA{}
\end{enumerate}

\end{enumerate}
\fi
\appendix

\section{Appendix}
\subsection{Score function estimator}\label{app:esti}
The empirical formulation of the score estimator from $N$ particles representing an unknown density $\rho(x)$ is
\begin{equation} \label{score_estim} 
    {\partial_\alpha} \ln \rho(x) \approx \sum_{i=1}^M \left( \sum^M_{k=1} B_{i k}(x)\sum^N_{l=1} 
\nabla_{X_l} K(X_l, Z_k)  \right),
\end{equation}

with $B_{ik}$ denoting the $i$-th row, and  $k$-th column of the matrix $B(x)$ defined as
\begin{equation} \label{eq:Ascore}
B(x) \doteq K(x,\mathcal{Z}) \left[ \lambda  I + 
(K(\mathcal{Z},\mathcal{Z}))^{-1} (K(\mathcal{X},\mathcal{Z}))^\top (K(\mathcal{X},\mathcal{Z}))\right]^{-1}(K(\mathcal{Z},\mathcal{Z}))^{-1} ,
\end{equation} 
where $\mathcal{X}=\{ X_i \}^N_{i=1}$ and $\mathcal{Z}=\{ Z_i \}^M_{i=1}$ denote the sets of samples and inducing points respectively, while $I$ stands for an $M \times M$ identity matrix. (Here we set the regularising constant $\lambda= 10^{-3}$).
We used an gaussian kernel
\begin{equation}
K(x,x') = \exp\left[-\frac{1}{2 {l}^2}\| x- x'\|^2\right],
\end{equation}
where the lengthscale $l$ was set to two times the standard deviation of the particle ensemble for each time step.

\subsection{Implementation details} \label{implementation}

Here we provide the algorithm for computing optimal interventions $u^*(x,t)$. 
Since the initial conditions for the flows $\rho_t(x)$ and $\tilde{q}_t(x)$ are delta functions centered around the initial and target state, $x_0$ and $x_1$, i.e $\rho_0(x) = \delta(x-x_0)$ and $\tilde{q}_t(x)=\delta(x-x_1)$, we employ a single stochastic step at the beginning of each (forward and time-reversed) flow propagation. 
Since the inducing point number $M$ employed in the gradient--log--density estimation is considerably smaller than sample number $N$, i.e.,~$M\ll N$, the overall computational complexity of a \emph{single} gradient-log-density evaluation amounts to $\mathcal{O}\left( N\,M^2 \right)$. 
We perform Euler integration for the ODEs, and Euler-Maruyama for stochastic simulations. For all numerical integrations we employ $dt= 10^{-3}$ discretisation step.
\if False
\renewcommand\thealgocf{A\arabic{algocf}}
\vspace{12pt}
\begin{algorithm}[H]\label{alg1}
\SetAlgoLined
\DontPrintSemicolon
\KwIn{
    $N,M$: scalars, number of particles and number of inducing points\\
    \hspace*{35pt}$t_0,t_1, dt$: scalars, initial and final timepoints, and discretisation step\\
    \hspace*{35pt}$x_0, x_1$: $1 \times d$, $1 \times d$ initial and target state \\
    \hspace*{35pt}$f$: drift function\\
    \hspace*{35pt}$\sigma$: noise amplitude\\
    \hspace*{35pt}$U(x,t)$: function, path constraint (optional)\\
 }

\KwOut{$Z, B$: $d \times N\times (t_1-t_0)/dt$, samples from forward flows $\rho_t(x)$ and $q_t(x)$\\
\hspace*{45pt}$u^*(x,t)$: functions from $\mathbb{R}^d \rightarrow \mathbb{R}^d$ for each $(t_1-t0)/dt$ time step, time- and \hspace*{85pt} state-dependent controls }
\BlankLine

$k = (t_1-t_0)/dt$   \tcp*{number of timesteps,}

\tcp*{,Forward propagation of flow $\rho_t(x)$             \hspace{183pt}               }

$Z_{ti=0} = x_0$   \tcp*{initialise particles' positions, }

$Z_{ti=1} = Z_{0} + dt f(Z_{0},t_0) + \sigma \mathcal{N}(0,\sqrt{dt})$   \tcp*{1st step is stochastic, }
For $ti= 2:k$ \tcp*{deterministic propagation , }

 \hspace{36pt}$Z_{ti+1} = Z_{ti} + dt \left( f(Z_{ti},t) -\frac{1}{2} \sigma^2 {\color{cadetblue}{\nabla \log \rho(Z_{ti})}}  \right) $   

\hspace{36pt} If $\exists$ path cost $U(x,t)$: \\

\hspace{66pt} $W = \exp{\left(-U(Z_{ti+1},t) \,dt \right) }$ \\
\hspace{66pt} $T^* =$ EnsembleTransformParticleFilter$(Z_{ti+1},W)$\\
\hspace{66pt} $Z_{ti+1} = Z_{ti+1} \cdot T^*$\\

\tcp*{,Time-reversed propagation of flow $q_t(x)$             \hspace{155pt}  }
$B_{ti=k} = x_1$   \tcp*{initialise particles' positions, }

\tcp*{1st step is stochastic }

$B_{ti=k-1} = B_{k} - dt \left( f(B_{k},t_1)    +\frac{1}{2} \sigma^2 {\textcolor{cadetblue}{\nabla \log \rho(Z_{k})}} \right)+ \sigma \mathcal{N}(0,\sqrt{dt})$   

For $ti= k-2:0$ \tcp*{deterministic propagation , }
\hspace{36pt} $B_{ti-1} = B_{ti} + dt \left( f(B_{ti},t) +\frac{1}{2} \sigma^2 {\color{cadetblue}{\nabla \log \rho(Z_{ti})}} -\frac{1}{2} \sigma^2 {\color{cadetblue}{\nabla \log q(B_{ti})}}  \right) $

\tcp*{,Compute $u^*(x\,t)$             \hspace{265pt}  }
For $ti= 2:k$

\hspace{36pt} $u^*(x,ti) = \sigma^2 {\color{cadetblue}{\nabla \log q(B_{ti})}} - \sigma^2 {\color{cadetblue}{\nabla \log \rho(Z_{ti})}} $

\caption{{Deterministic Particle Flow (DPF) control} }
\end{algorithm}

\fi
For the numerical experiments with path constraints, we solved the optimal transport problem with the implementation 
of FastEMD~\cite{pele2009}.

For some of the visualisations of our results we used the Seaborn~\cite{waskom2021seaborn} python toolbox.

\if False
\subsection{Controlling evolving populations} \label{evol}

For an evolving population, the main evolutionary drivers comprise fitness and mutation forces that continuously adjust the composition of phenotypes within the population, while genetic drift perturbs the whole process stochastically. 
We describe the evolution of the \emph{mean phenotypes} $\mathbf{x}:=(x,y)$ of the population by the overdamped Langevin equation 
\begin{equation} \label{eq:pheno1}
    d\mathbf{x} = C \cdot\nabla F(\mathbf{x}) dt + \sigma d{W},
\end{equation}
with $F(x)$ denoting the phenotypic landscape in the presence of natural selection~\cite{provine1989sewall}, where the landscape axes represent different phenotypic traits,
and $\sigma$ the noise amplitude that rescales the genetic drift, i.e. the stochastic term, according to the population size $n$ and the covariance matrix $C$, $\sigma = C^{1/2} n^{-1}$. The gradient of the landscape $ f(x) = C \cdot \nabla F(x)$ captures the adaptive pressures under natural selection.

Here, we assume an asymmetric rugged landscape~\cite{barton2007evolution}, that may arise in small sized populations with small variance, and multi-modal individual fitness functions ~\cite{whitlock1995variance}, supported also by empirical findings indicating asymmetry in selection landscapes~\cite{kingsolver2001strength}.
In this setting, optimal control can be thought of as artificial selection that diverts evolving populations from strictly following natural selection, and drives them to externally imposed target states~\cite{nourmohammad2021optimal}.
Path constraints are essential to prevent changes of co-varying phenotypic traits. Artificial selection to promote the prevalence of a phenotypic trait in a population may lead to undesired variations along covarying traits. Therefore, path constraints may be employed to reduce the fluctuations along covarying, but not targeted traits. For simplicity we consider the covariance matrix $C=I$ constant, given the much smaller timescales upon which its fluctuations unfold~\cite{nourmohammad2013evolution}, and its weak dependency on the evolutionary selection strength~\cite{held2014adaptive}.

The dynamics of Eq.(\ref{eq:pheno1}) describe the evolution of populations in the presence of natural selection towards an evolutionary optimum, captured by the maximum of the adaptive landscape, adhering thereby to physiological and environmental constraints. 

To study and understand the outcomes and dynamics of adaptive evolution, we need to devise intervention protocols that drive phenotypes towards non-evolutionary optimum states $\mathbf{x^*}$, or through evolutionary trajectories that deviate from the gradient of the phenotypic landscape. This intervention is implemented through artificial selection, which, following~\cite{nourmohammad2021optimal}, we formulate here as a time- and state- dependent perturbation $\mathbf{u}(x,t)$ to the natural selection
\begin{equation}
     d\mathbf{x} = \Big( f(\mathbf{x}) + u(\mathbf{x},t)\Big)  dt + \sigma d\mathbf{W}.
\end{equation}

\fi

\end{document}